# 40 Years of SCES at Los Alamos

Z. Fisk, J. L. Smith and J. D. Thompson*
Los Alamos National Laboratory, Los Alamos, NM 87545, USA

*Corresponding author: jdt@lanl.gov

Abstract: Reports of unconventional superconductivity in $UBe_{13}$ in 1983 and soon thereafter of the possible coexistence of bulk superconductivity and spin fluctuations in $UPt_3$ marked the beginning of a 40-year adventure in the study of strongly correlated quantum materials and phenomena at Los Alamos. The subsequent discovery and exploration of heavy-fermion magnetism, cuprates, Kondo insulators, Ce- and Pu-115 superconductors and, more broadly, quantum states of narrow-band systems provided challenges for the next 30 years. Progress was not made in a vacuum but benefitted from significant advances in the Americas, Asia and Europe as well as from essential collaborations, visitors and Los Alamos students and postdocs, many subsequently setting their own course in SCES. As often the case, serendipity played a role in shaping this history.

## I. INTRODUCTION

Strongly correlated electron systems (SCES) as a field of study has a rich history, and we recount a small part of that history in the Americas. Our primary aim, though, is to share some memories of how the study of heavy-fermion materials evolved at Los Alamos. Before coming to that history, we begin with some earlier context. Already by the mid-1970's, the theoretical and experimental study of SCES was emerging as an active field, especially in Europe, Japan and the U.S but also in South America and India. In the Fall of 1976, Ron Parks organized what reasonably should be considered the first international SCES conference,[1] certainly in the Americas, that brought together early pioneers including Blas Alascio, Jim Allen, Phil Anderson, Merwyn Brodsky, Kurt Buschow, Bernard Coqblin, Seb Doniach, Duncan Haldane, Lester Hirst, Tadao Kasuya, Jon Lawrence, Brian Maple, Hans Ott, Peter Riseborough, Frank Steglich, John Wilkins, Dieter Wohlleben and Chandra Varma, just to mention a few. Invigorated by recent discoveries, conference attendees discussed and debated the origin and interpretation of mixed valence, charge and spin fluctuations and implications of the Anderson model, the Kondo lattice, Doniach's recently proposed phase diagram, and stability of an emergent heavy Fermi-liquid state at low temperatures. These remain core subjects of SCES conferences even today. There was no representation of Los Alamos work at Parks' conference.

At the time of this conference, paramagnetic CeAl$_3$ had been discovered to be the first example of a material with very massive charge carriers [2] and NpSn$_3$ to be what now would be called a heavy-fermion antiferromagnet (M. B. Brodsky in ref. [1]), but heavy-fermion superconductivity would not be reported for another three years.[3] In hindsight, the possibility of heavy-fermion superconductivity
might have been envisioned even earlier. In 1975, Bucher et al. [4] noted that the electrical resistivity of UBe$_{13}$ dropped to zero at 0.97 K and that its specific heat grew to a very large value at 1.8K, the lowest temperature of these measurements. Despite only a weak depression of the transition temperature by a magnetic field and a large diamagnetic response persisting even when the sample was ground into a
powder, these authors concluded that evidence for superconductivity was extrinsic. At least two factors likely influenced this conclusion. Evidence for pair-breaking by magnetic (Kondo) impurities in conventional superconductors, from the work of Brian Maple and others,[5] as well as the established theory of magnetic pair-breaking by Abrikosov and Gorkov[6] pointed to the unlikely possibility that superconductivity could develop in UBe$_{13}$ which exhibited local-moment-like magnetic susceptibility. In addition, there was the correlation between ground states and R-R spacing, where R =Ce, U, Np and Pu, discovered by Hill at Los Alamos.[7] According to this correlation, the very large U-U spacing in UBe$_{13}$ should put it deep into the magnetically ordered regime and not that occupied by superconductors at small U-U spacing. A subsequent stroke of insight would upend prevailing views of the relationship between magnetism and superconductivity.

As an outgrowth of Bernd Matthias' earlier discovery of itinerant ferromagnetism in ZrZn$_2$, in the late 1970's, work at Los Alamos [8] had discovered the first "itinerant antiferromagnet" TiBe$_2$, a conclusion based on a comparison of theoretical predictions by Enz and Matthias [9] and magnetic susceptibility measurements [8]. A year later, Greg Stewart found an upturn in $C/T$ below ~15 K [10], similar to behavior found earlier in the spin-fluctuation compound UAl$_2$ where $C/T$ at low temperatures increased as $T^2 \ln T/T_{sf}$. [11] The specific heat and resistivity of UAl$_2$ and related materials also had drawn the interest of two young theorists,[12] J. R. Iglesia-Sicardi and A. A. Gomes, who, after working with Coqblin and Jullien in Orsay, returned to positions at the Universidad Nacional del Sur in Argentina and at Centro Brasileiro de Pesquisas in Rio de Janeiro, respectively. A lively nucleus of scientists interested in SCES was forming in South America. Carlos Rettori brought new expertise in electron-spin resonance studies of the Kondo effect to the University of Buenos Aires [13] and soon thereafter to the Universidade Estadual de Campinas [14]. Theorists Mucio Continentino and Gaston Barberis joined Gomes in Brazil. A couple of years later, Elisa Baggio-Saitovitch would establish Mossbauer spectroscopy in Rio de Janeiro. In Argentina, Blas Alascio, a theorist at Centro Atomico Bariloche, was interacting with experimentalists Julian Sereni as well as Francisco (Paco) de la Cruz, Maria Elena de la Cruz and Ana Celia Mota, who were key in developing low-temperature capabilities that were essential to SCES studies there.

## II. A TIME OF CHANGE

Like the report of a resistive transition in UBe$_{13}$, the discovery by Frank Steglich and coworkers of superconductivity that developed out of a normal state of "heavy quasiparticles" in strongly paramagnetic CeCu$_2$Si$_2$ [3] initially attracted little attention, being cited only 3 times in 1980, 6 times in 1981 and 9 times in 1982.[15] In spite of the retrospectively historical importance of this discovery for changing the direction of SCES research, sample dependence [16] discouraged some from acknowledging its importance but so did the community's reluctance to accept the possibility of superconductivity in a strongly paramagnetic metal. A different kind of change also was taking place at Los Alamos. In 1980, an outcome of the passing of Matthias, who had had a significant influence on the direction of materials, especially actinide, research throughout Los Alamos but who also was totally uninterested in the Kondo physics, was that Fisk (ZF) joined efforts of Smith (JLS) and Stewart in a team in a chemistry and metallurgy division. Together, ZF and JLS continued to explore properties of TiBe$_2$ and its alloys as well as systematics of magnetism in transuranics.[17] During a visit to Los Alamos in the summer of 1982, Jon Lawrence participated in discussions of these transuranic systematics but also brought a problem – critical behavior at the valence instability in Ce$_{1-x}$Th$_x$. Instead of accessing critical behavior with Th doping, Lawrence proposed that a pressure capability recently developed by Thompson (JDT) in a physics division might be an alternate route. His proposal led to the first collaboration among us and experimental support [18] for a Kondo-volume collapse model [19] of the $\gamma$-$\alpha$ transition in Ce. This project, which began a continuing collaboration with Lawrence, was the first work on Kondo physics at Los Alamos since Bill Steyert and visitor Melvin Daybell had made early studies of the Kondo-impurity effect over 15 years earlier in the physics group.[20]

In the Fall of 1982, Hans Ott corresponded with ZF about his suspicion that UBe$_{13}$ "might show equivalent properties to CeCu$_2$Si$_2$." A letter received by JLS from Ott in November expressed these suspicions and requested polycrystalline samples, initially to test his conjecture as soon as possible. But, Ott and ZF were well aware of the sample dependence of CeCu$_2$Si$_2$ properties and the need to validate polycrystalline results in single crystals. Poly- and single crystals were prepared immediately and shipped to Ott at ETHZ. Ott measured specific heat and the Los Alamos team measured ac susceptibility and resistivity. With reproducibility of superconductivity in polycrystalline samples and single crystals, they submitted their results to Physical Review Letters in March, 1983 and the paper was published two months later.[21] Not only did this work demonstrate that superconductivity in UBe$_{13}$ was intrinsic and unconventional, it also lent credence to heavy-fermion superconductivity in CeCu$_2$Si$_2$. Nevertheless, there still was some lingering possibility that these two examples were just quirks of Nature. These doubts were short-lived.

Also in the Fall of 1982, Jaap Franse at Amsterdam sent to Los Alamos a collection of papers by his group, including one that would be published by P. H. Frings et al. on magnetic properties of U$_x$Pt$_y$ compounds at temperatures above 1.4 K.[22] Specific heat and magnetic susceptibility/magnetization of one of those compounds, UPt$_3$, were similar to those of TiBe$_2$ and UAl$_2$. During a subsequent visit to the Amsterdam group, Stewart asked for and was given a piece of Czochralski-grown UPt$_3$ so that he could study the specific heat in more detail. By the time Stewart returned in June of 1983, and independently of those discussions, ZF already had flux-grown, high-quality single-crystal whiskers of UPt$_3$ and measured their resistivity to $^4$He temperatures where $\rho$(T) still was dropping. To avoid a possible conflict, Stewart explained to Franse that he should study the crystals grown by ZF and not the Czochralski piece from Amsterdam; Franse agreed. Interested to see how much lower $\rho$(T) might drop, ZF asked Jeff Willis, also a member of his team, to extend resistivity measurements to dilution-fridge temperatures. To everyone's surprise, the resistivity dropped to zero at 0.5 K. After quickly modifying a dilution fridge to enable specific heat measurements, Stewart and Willis confirmed that the resistive transition was to a bulk superconducting state.[23] By the time their paper was submitted to Phys. Rev. Lett. in October 1983, Franse's group had independently found a resistive transition in their Czochralski-grown crystals and subsequently would confirm the Los Alamos results.[24] The Los Alamos submission to Phys. Rev. Lett. did not go smoothly, initially being rejected in part because of a somewhat rounded specific heat anomaly at $T_c$—later shown to be due to two nearby anomalies. A second round of reviews included the comment "I hope the authors do not continue to risk their significant scientific reputations and push for publication of this work." Not being discouraged, the authors appealed the decision and the paper was accepted four months after its initial submission. Now with three examples, there was no doubt that heavy-fermion superconductivity was real and likely mediated by spin fluctuations, and the community responded with a burst of theoretical and experimental activity. 1983 was a defining year that marked the beginning of what would become a 40-year adventure in the study of strongly correlated quantum materials and phenomena at Los Alamos and beyond.

By the end of 1984, the Los Alamos team had discovered heavy-fermion paramagnetism in CeCu$_6$ and antiferromagnetism in NpBe$_{13}$, U$_2$Zn$_{17}$ and UCd$_{11}$ [25] and supplied crystals of these as well as UBe$_{13}$ and UPt$_3$ to collaborators around the world. Neutron scattering, optical spectroscopy, photoemission, ultrasound, and NMR, which were not available then at Los Alamos, would be especially informative. Canadian neutron scatterer Bill Buyers and spectroscopist Tom Timusk were some of those early collaborators as were experimentalists at Bell Labs and many academics, especially those associated with various campuses of the University of California. To a substantial extent, experimental activity was driven by new materials and exploration of their properties, but theory had to come to grips with the origin of unconventional superconductivity and of the heavy-fermion state that developed in a lattice of 'Kondo impurities.' Early on, Chandra Varma [26] and many others tackled these issues from both phenomenological and microscopic perspectives. The rapid pace of experimental and theoretical advances was, in no small part, due to Los Alamos acting as a snail-mail predecessor of the arXiv. Preprints flowed in; JLS forwarded collections of those and Los Alamos preprints to over 450 recipients on a monthly basis. In these and the coming few years, he mailed around $10^7$ pages of pre- and post-publications.

Publications by us in these early days often carried the by-line 'Center for Materials Science,' though only JLS was officially a member (Chairman) of this new organization with one of its missions being to host distinguished university and industrial scientists for collaboration with Lab staff.  Heavy-fermion research was a logical focus of outreach objectives, and the Center sponsored visits by many of the world's leading SCES scientists, especially theorists. An outgrowth of so-called summer working groups at the Center was a critical assessment by some of those visitors of the state of heavy-fermion theory [27] and somewhat later a broader review of heavy-fermion materials and their understanding [28] that was shaped, in part, by discussions with visitors Gabe Aeppli, Bertram Batlogg, Hans Ott, Tom Rosenbaum, Doug MacLaughlin and Brian Maple. The Center also enabled extended visits of Ron Parks in 1984 and 1985 to collaborate with JDT in their study of the pressure-response of Kondo-lattice systems and more particularly how this response might reflect the competition between Kondo and RKKY interactions embedded in Doniach's phase diagram.[1]  A consequence of this collaboration was the anticipation of pressure-induced antiferromagnetic quantum-critical points near 0.9 GPa in $CeRh_2Si_2$ and somewhat above 1.8 GPa in $CePd_2Si_2$.[29]  Unfortunately, this fruitful collaboration was cut short with Parks' passing in April 1986. By the end of 1985, Stewart had left Los Alamos to start his own SCES group in Florida, JLS had taken on more time-consuming responsibilities at the Center, and ZF had joined JDT in the physics group. Nevertheless, efforts to discover new materials continued, with reports of heavy-fermion behavior in $YbAgCu_4$ [30] and a family of ternary compounds derived from $CeCu_5$.[31] With the growth of new examples and a new appreciation that previously known materials fall in a continuum of fermion heaviness, correlations among them began to emerge, for example, a Wilson ratio $\chi(0)/\gamma(0)$ essentially equal to that of a non-interacting Fermi gas [27] and a common relationship between the ratio of a $T^2$ coefficient of resistivity and Sommerfeld coefficient, reported by Kadowaki and Woods [32] in Canada.

   Insights from the study of heavy-fermion materials set the stage for the SCES community to respond immediately to surprising reports of superconductivity in copper oxides.[33, 34]  Virtually all in the community immediately redirected much of its effort. Los Alamos was no exception. At the infamous 1987 APS March Meeting we reported for the first time the existence of superconductivity above 90 K when non-magnetic Y was replaced by a magnetic rare earth (RE) in what would become $REBa_2Cu_3O_7$. This discovery and the observation of RE ordering at much lower temperatures appeared in print soon thereafter. [35, 36] Even Edward Teller would stop by to ask what was new, what was understood and what wasn't, and Gene Wells, then Editor of Phys. Rev. Lett., spent 1986-1987 working in the lab with us [37].  Again, the Center for Materials Science provided focus and a certain 'convening authority' through a new working group led by Bob Schrieffer and that included David Pines, Doug Scalapino and Elihu Abrahams. Across the Americas and more broadly, those early days of high-$T_c$'s stimulated remarkable advances in appreciating and understanding the physics of strong correlations, heavy-fermion physics, clearer insights into the possibility of spin-mediated superconductivity and powerful new experimental techniques for probing SCES. Muon-spin spectroscopy at Canada's TRIUMF facility was one of those dual-use examples that came to prominence by benefiting the study of both cuprate and heavy-fermion physics.

After the initial world-wide push to understand existing and discover new cuprate superconductors, an equilibrium between heavy-fermion and cuprate research began to emerge at the end of this remarkable decade. This is reflected in presentations at the September 1989 International Conference on the Physics of Strongly Correlated Electron Systems, organized by Jack Crow, JLS and others from Los Alamos [38] in Santa Fe, and at the 6$^{th}$ International Conference on Valence Fluctuations [39], organized by Gaston Barberis and his colleagues in Rio de Janeiro during July 1990. The Rio conference, in particular, highlighted major advances in the study of heavy-fermion and Kondo problems being made in South America but also reflected remarkable progress on phenomenological and technical approaches to the two-Kondo-impurity/Anderson-Lattice models that are basic to heavy-fermion physics.

A major development at this time was the report by a Berkeley-Grenoble-Los Alamos collaboration of two bulk superconducting transitions in UPt$_3$.[40] This discovery provided strong support for an unconventional superconducting order parameter in UPt$_3$, an idea that also had been raised in the context of the earlier discovery of a second bulk transition below $T_c$ in Th-doped UBe$_{13}$[41]. Even today the nature of the second transition in (U,Th)Be$_{13}$ remains unclear, though it's likely to be to another superconducting phase.[42] At Los Alamos, exploration of Ce-Pt compounds grown out of a Bi-flux led to crystals of cubic Ce$_3$Bi$_4$Pt$_3$, which were not heavy-fermion metals but unexpectedly exhibited a small gap in electrical resistivity [43] in contrast to metallic La$_3$Bi$_4$Pt$_3$. As argued, the activated transport derived from hybridization between $4f$ and conduction electrons. Ce343, as it became known, was half-in-jest termed a 'Kondo insulator' [44], a half-filled Anderson lattice.[45] A natural question was whether it would be possible to grow Yb343 from Bi-flux. Instead of a 3-4-3 composition, single crystals with 1-1-1 composition in the half-Heusler structure appeared in growths. One of the first studied was YbBiPt which we found to have a very large Sommerfeld coefficient [46] and a spin-density transition near 0.4 K that could be suppressed to T=0 with a field of only 30 kOe [47]. Crystals of other family members RBiPt (R = Ce-Lu, except Pm and Eu) followed immediately [48] and eventually would regain attention due to the possibility of their hosting Weyl fermions.[49, 50]

### III.  THE NEXT THIRTY YEARS OF SCES

The next three decades began with the first conference in the Americas with the title Strongly Correlated Electron Systems, which was chaired by ZF, Pradep Kumar and Maple during the summer of 1993 in San Diego.[51] Though experimental and theoretical progress on cuprates and many of the known heavy-fermion/Kondo-lattice materials was well-represented, a relatively new theme of non-Fermi-liquids (NFL) received attention. Much of the experimental work came from Maple's group that had been studying M$_{1-x}$U$_x$Pd$_3$ (M=Y, Sc, Th, La) alloys as well as from Bohdan Andraka and Stewart (Univ. Florida) who also reported a logarithmically diverging specific heat divided by temperature in several other U- and Ce-based alloys. The origin of the NFL was not clear, possibly manifesting a marginal Fermi-liquid state in $UPd_3$ alloys through a 2-channel quadrupolar Kondo effect [52] or, as suggested by Andraka and Tsvelik [53], from fluctuations of an unknown (possibly antiferromagnetic) order parameter in the vicinity of a $T=0$ critical point. Just prior to [54] and at the conference, Andy Millis reported theoretically expected NFL signatures of critical fluctuations at a $T=0$ spin-density transition. This work, along with earlier theory by Hertz and Moriya, would become known as the 'conventional' HMM theory of quantum criticality in which only fluctuations of an order parameter are quantum-critical as a second-order transition is tuned to zero-temperature, a concept envisioned by Doniach in the 1976 conference organized by Parks. In parallel, Continentino derived generalized scaling properties of a system close to a quantum-phase transition. [55] Almost immediately, though, experiments questioned the applicability of the HMM framework to account for spin dynamics of the non-Fermi-liquids $UCu_{5-x}Pd_x$ (x=1, 1.5) in which the dynamical susceptibility scaled as energy/temperature ($\omega/T$). [56] Such scaling should not happen in the HMM theory. Exploring the ideas of non-Fermi-liquids and quantum criticality would become a theme of SCES research.

In spite of joining Florida State University and the newly established National High Magnetic Field Laboratory in 1994, ZF continued active involvement in heavy-fermion work at Los Alamos during extended summer visits and whenever time would allow. In his absence, we returned to $CeRh_2Si_2$, which our earlier work had suggested to have a quantum critical point below 1 GPa, and found pressure-induced superconductivity with a maximum $T_c$ of 0.4 K near the $T=0$ antiferromagnetic/paramagnetic boundary at ~0.8 GPa. [57] This discovery, in light of earlier work on $CeCu_2Ge_2$ [58] and independent of a simultaneous report of pressure-induced superconductivity in $CePd_2Si_2$ by a Cambridge group [59], illustrated that heavy-fermion superconductivity 'liked' to emerge at an antiferromagnetic QCP. Over the years, the idea that fluctuations around a QCP might provide an attractive pair interaction has influenced, with some success, the search for new examples of unconventional superconductivity in SCES.[60]

Though we did not appreciate it, the late spring/early summer of 1997 could have been a turning point in research direction. In the course of exploring Bi-, Ga-, In- and Sn-rich Ce-based ternaries and quaternaries, ZF found crystals of $CeRhIn_5$ and soon thereafter crystals of $CeIrIn_5$ in some In-rich growths. Though ZF deduced that these materials were tetragonal from x-ray measurements on Gd-analogs, it was postdoc Evagelia Moshopoulou who eventually would solve the crystal structure. [61] Besides all the many new materials produced by ZF that summer and on-going studies of cuprates and manganites, John Sarrao, who had (re)joined the SCES group earlier in 1997 as our 'in-house' crystal grower, was interested in a family of $YbXCu_4$ materials, among others. One of those materials was atomically ordered crystals of $YbInCu_4$ that grew out of a flux in which ZF earlier had made a mistake in composition. These high-quality crystals showed a sharp isomorphic transition near $T_v \approx 40$ K that was accompanied by a pronounced change in the valence of Yb [62], similar to what happens at the first-order isostructural volume-collapse transition in Th-doped Ce [16]. In both cases $T_v$ could be suppressed to zero in a sufficiently high magnetic field,[63,64] with the phase boundary $T_v(B)$ being described quantitatively by a simple entropy argument that relied solely on the experimentally justified assumption that the characteristic energy scale of the phase at $T > T_v$ is much smaller than that of the phase at $T < T_v$ [63]. These YbXCu$_4$ and other materials consumed bandwidth of our basic characterization capabilities that had grown to include photoemission (Al Arko), NMR (Masashi Takigawa/Chris Hammel/Nick Curro), µSR (Bob Heffner), neutron diffraction (George Kwei/Wei Bao), a range of transport and thermodynamic measurements above (Mike Hundley) and below 1 K (Roman Movshovich), resonant ultrasound (Al Migliori), ultrafast spectroscopy (Toni Taylor) and initial members of the newly established National High Magnetic Field Pulsed Field Facility. Even with this larger contingent, almost a full year later all we knew was that CeRhIn$_5$ ordered antiferromagnetically with about 0.3$R$ln2 entropy below $T_N = 3.8$ K and that CeIrIn$_5$ had a bulk phase transition near 0.4 K even though its resistance dropped to zero around 1 K. Similar to the earlier affair with UBe$_{13}$, we were blinded by the facts, concluding erroneously that in CeIrIn$_5$ the resistance drop came from a minor second phase and the specific anomaly near 0.4 K was due to antiferromagnetic order that developed out of a heavy-fermion state—as we had found in YbBiPt. Activity picked up in the summer of 1998 when ZF's student Cedomir Petrovic arrived in Los Alamos to continue his PhD research. A part of that research was to determine if the resistive transition in CeIrIn$_5$ was extrinsic, as we expected. Over the course of several months, he polished and etched many crystals of CeIrIn$_5$ to see if the 'second phase' and hence resistivity transition could be removed, a test of patience all to no avail. From our Doniach phase-diagram studies and the relatively small entropy below $T_N$ in CeRhIn$_5$, we speculated that magnetic order might be suppressed with relatively modest pressure, and postdoc Helmut Hegger began to pursue that idea in late 1998. Finally, two years after first having crystals of CeIrIn$_5$ and CeRhIn$_5$, Petrovic and Movshovich had discovered evidence from *ac* susceptibility and specific heat measurements for bulk superconductivity at 0.4 K in CeIrIn$_5$ and that the anisotropic critical magnetic fields of the bulk and resistive superconducting transitions scaled [65] as well as that replacing Ir with Rh increased the bulk $T_c$ [66]; Hegger had discovered pressure-induced superconductivity in CeRhIn$_5$ with a maximum $T_c$ of over 2K, by far the highest $T_c$ of any known heavy-fermion superconductor.[67] We attributed the high $T_c$ to the quasi-two-dimensional structure of CeRhIn$_5$. Before Petrovic returned to Florida State at the end of 1999, he also discovered resistive evidence for pressure-induced superconductivity in Ce$_2$RhIn$_8$, a more three-dimensional variant of CeRhIn$_5$.[68]

It had been frustrating not being able to grow crystals of CeCoIn$_5$, which would have been a logical extension in the sequence Ir-Rh-Co, especially in light of knowing how to grow the Rh and Ir materials and that much earlier during his PhD studies Yuri Grin had discovered Ga-analogs with heavier rare-earth elements (R) in the R$_2$CoGa$_8$ and RCoGa$_5$ homologous series [69]. While writing his PhD thesis, though, Petrovic tried substituting Ir with Co in CeIr$_{1-x}$Co$_x$In$_5$ and in the course of optimizing crystal-growth conditions discovered how to grow CeCoIn$_5$. Fellow graduate student Fivos Drymiotis determined from SQUID magnetometry that the crystal became superconducting around 2 K. These events occurred while ZF was traveling, and he learned of them through a fax from Petrovic once he arrived at Los Alamos. The Los Alamos group soon applied all capabilities to explore basic superconducting and normal-state properties of this new material that by the summer of 2000 was part of a larger family $Ce_nT_mIn_{3n+2m}$, where T=Rh or Ir, $n$=1 or 2, and $m$=1 that we had prepared, begun to study and reported initial results on (including $CeCoIn_5$ in the oral presentation) at the joint ICM/SCES conference during August, 2000 in Recife, Brazil[70]. Much work on these materials by Los Alamos and other groups followed.

Before the end of June, 2000, we had submitted a $CeCoIn_5$ manuscript where we noted that, as in the cuprates, $T_c$ reached ~20% of the relevant temperature scale $T_{sf}$ for magnetically mediated superconductivity, an important insight provided by Phillipe Monthoux.[71] This led to our speculating that $T_c$ might be even higher in a d-electron analog of $CeCoIn_5$ in which $T_{sf}$ might be higher. This speculation would become reality a few months later. Perhaps auspiciously but coincidentally, the influential review "How do Fermi liquids get heavy and die?" by Piers Coleman, Catherine Pepin, Qimiao Si and Revas Ramazashvili appeared later in the same volume [72] where these authors discussed experimental observations in the context of various theoretical approaches to quantum criticality, including a new mechanism that involved a break-down of the composite nature of heavy electrons, an associated jump in the Hall number and $\omega/T$ scaling of the dynamical susceptibility (which also is characteristic of a marginal Fermi liquid [73]).

By August, 2001, there was strong evidence that superconductivity in $CeIrIn_5$ and $CeCoIn_5$ was d-wave [74] and as also reported by Yoshichika Onuki et al. at the SCES conference [75] in Ann Arbor (organized by Meigan Aronson and Jim Allen) that the electronic structure of $CeRhIn_5$ and $CeCoIn_5$ was quasi-two dimensional.[76] Soon after SCES2001, Sarrao was trying to grow crystals of $PuGa_3$ and added a bit of Co to the mix, believing that it would lead to the growth of large crystals. Large crystals did form and JDT measured their susceptibility in Dec., 2001. Much to our surprise, nearly perfect diamagnetism developed below 18.5 K. It took weeks to find that the large crystals were, in fact, $PuCoGa_5$ with the same structure as the Ce-based 115s. (From x-ray measurements, we believed initially that the compound was $PuCo_2Ga_4$. Even specific heat measurements in May, 2002 assumed this composition.) Gerry Lander happened to be visiting Los Alamos then, and we told him about these results, which he relayed to his colleagues at the Institute for Transuranics in Karlsruhe. They extracted small crystals from arc-melted material and confirmed superconductivity through resistivity and susceptibility measurements. With this confirmation, a manuscript on superconductivity in $PuCoGa_5$ was accepted without difficulty.[77]

The enhanced but smaller Sommerfeld coefficient of $PuCoGa_5$ compared to $CeCoIn_5$ implied a roughly order of magnitude larger $T_{sf}$ in the Pu-based material whose $T_c$ also was roughly an order of magnitude higher. This is the trend we had speculated should be present for magnetically mediated superconductivity and is consistent with the greater spatial extent, and hence *f-c* hybridization, of Pu 5*f* wavefunctions. In just two years after its discovery, Curro used NMR to show that the superconductivity in $PuCoGa_5$ was unconventional, likely mediated by antiferromagnetic spin fluctuations, and that its properties followed a common relationship between $T_c$ and $T_{sf}$ which spanned more than two orders of magnitude, from $CeCu_2Si_2$ to the Hg- and Tl-based cuprates, in superconductors believed to be magnetically mediated.[78] Such as correlation had been proposed earlier by Moriya and Ueda.[79]

The Ce- and Pu-based 115s and subsequently discovered related compounds consumed much attention of the SCES effort at Los Alamos and attracted broad interest, but these materials were only part of a burst of activity that marked the first decade in the 21st century. In Canada, Louis Taillefer, who was first to discover quantum oscillations in $UPt_3$ with his advisor Gil Lonzarich,[80] had joined the Univ. of Toronto in 1998. Though his interests focused more on the normal state of cuprates, he subsequently established a fruitful collaboration with Petrovic to study the unusual superconducting and normal state properties of the Ce115s. After Taillefer moved to Sherbrooke, Stephen Julian returned to the Univ. of Toronto, also by way of the Cavendish, where he established high pressure and very low temperature capabilities that allowed detailed quantum-oscillation studies, particularly in heavy-fermion materials such as $UPt_3$, $YbRh_2Si_2$ and $CeRu_2Si_2$ but also SCES more broadly. Individually and through their association with the Canadian Institute for Advanced Research Quantum Materials group, both played key roles in raising the visibility of strongly correlated materials and phenomena in Canada and the US. Some years later, former Los Alamos postdocs Jeff Sonier (to Simon Fraser), Andrea Bianchi (to Univ. Montreal), Sarah Dunsinger (to TRIUMF/Simon Fraser) and Meigan Aronson (to Univ. British Columbia) would join the increasingly active SCES efforts in Canada. To the south, Pascoal Pagliuso and Ricardo Urbano, who also had been postdocs at Los Alamos, joined theorist Eduardo Miranda and colleagues in Campinas to pursue studies of the 115s and NFL behavior. They, along with Alvaro Ferraz and others, organized workshops on strange metals, quantum criticality and topology, and unconventional superconductivity in Brasilia and Natal that featured state-of-the-art developments in the study of SCES, while Elisa Saitovitch and Mucio Continentino organized SCES 2008 in beautiful Buzios [81]. SCES 2008 continued the exciting developments reported during SCES 2007 [82], which had been organized in Houston by Qimiao Si and Paul Chu. Like earlier and future SCES conferences as well as numerous other correlated electron workshops, these conferences received support from the Institute for Complex Adaptive Matter/International Institute for Complex Adaptive Matter, a virtual institute founded at Los Alamos in 1998 by Pines and ZF to promote the study of complexity and its commonality in systems ranging from biological to heavy-fermion. In Argentina, Julian Sereni and collaborators (especially from Europe) were active studying a broad range of correlated magnets and paramagnets as well as organizing SCES workshops in Bariloche that brought together leaders of the SCES community from around the world. SCES in the Americas was strong; nevertheless, many outstanding questions remained, perhaps most notably being how does a heavy-fermion band develop at low-temperatures out of a lattice of local moments at high temperatures. The continued discovery of new heavy-fermion materials and phenomena allowed at least a phenomenological picture of that process.[83]

Of many open questions posed by SCES, one of interest to Los Alamos was the relationship among magnetic order, quantum criticality and unconventional superconductivity. CeRhIn$_5$ provided an exemplary case where its antiferromagnetic transition ($T_N$) could be tuned toward $T$=0, but a dome of pressure-induced superconductivity prevented following $T_N(P)$ to a quantum-critical point. A maximum $T_c$ of pressure-induced superconductivity being much higher in CeRhIn$_5$ than in other examples opened the possibility of tracking the evolution of magnetic order to much lower temperatures as superconductivity was suppressed by an applied magnetic field. *AC* specific heat measurements revealed a line of field-induced quantum-critical points, inside the superconducting phase, [84] that terminated at a pressure where, in the normal state, deHaas-vanAlphen frequencies jumped to larger values found in CeCoIn$_5$. [85] Superconductivity in CeCoIn$_5$ itself developed out of a non-Fermi-liquid normal state [86, 87] and would be shown to host spin-density order inside its low-temperature, high-field superconducting state.[88] Initial and subsequent measurements on CeRhIn$_5$ [89] suggested that the quantum criticality was of an unconventional, Kondo-breakdown type [90, 72]. Further support for these ideas and their generalization to a 'global' quantum-critical phase diagram [91] would come from pressure-dependent thermopower measurements on Ir-doped CeRhIn$_5$ [92] and were part of lively discussions at SCES 2010 [93] which was organized in Santa Fe, NM by Sarrao and JDT.

An interesting new direction for SCES emerged in 2010 with the theoretical prediction that Kondo insulators, particularly SmB$_6$, might host a topologically protected surface state.[94] A flurry of worldwide activity ensued, with many experiments performed on crystals grown by ZF (now at U. C. Irvine) and his postdoc Priscila Rosa; Rosa soon would join JDT and Eric Bauer who now was leading the SCES-materials discovery effort at Los Alamos [95, 96]. Though electrical transport, angle-resolved photoemission and tunneling spectroscopies pointed to a conducting surface state, observations by the Univ. Michigan group of deHaas-vanAlphen oscillations that were consistent with a metallic surface in flux-grown (by ZF) SmB$_6$ still came as a surprise.[97] Even more surprising was a subsequent claim by a different group of quantum oscillations coming from the bulk of float-zone-grown SmB$_6$, i.e., there was a three-dimensional metallic Fermi surface in a material whose bulk by all other measures should be insulating.[98] This claim raised a few eyebrows and still is not resolved fully. At least in flux-grown crystals, dHvA experiments by a new member of the Los Alamos SCES effort, Sean Thomas, argued that quantum oscillations likely arose from aluminum (flux) inclusions and were not intrinsic to SmB$_6$.[99] Irrespective and independent of possible quantum oscillations, work at Los Alamos [100] and elsewhere [101] pointed to a conducting surface with massive charge carriers in SmB$_6$, a signature of strong electronic correlations. Progress on topological states and strong correlations was well-represented at SCES 2018, organized by Stephen Julian and part of the 21$^{st}$ International Conference on Magnetism held in San Francisco.[102]

The 2019 report by a NIST/Univ. of Maryland team of unconventional superconductivity in nearly ferromagnetic UTe$_2$ initiated a new direction for the SCES community.[103] Its unusually large and anisotropic upper critical field, a power-law dependence of specific heat and lack of a change in Knight shift below $T_c$ ~1.6 K suggested spin-triplet superconductivity that is potentially topological and a host of

Majorana fermions. A peculiarity of the superconductivity, however, was a large residual Sommerfeld coefficient of specific heat in the limit $T \to 0$ (~50 mJ/molK$^2$) that was nearly 50 % of $C/T$ above $T_c$ and that was argued to be an intrinsic feature. Support for these conclusions [103] soon followed with evidence for time-reversal symmetry breaking (TRSB) from Kerr rotation experiments on crystals that exhibited two bulk superconducting transitions in specific heat, just as had been found much earlier in UPt$_3$. [104] These results placed constraints on the symmetry of possible multi-component order parameters and suggested the 'likely' possibility of Weyl points. Worldwide theoretical and experimental activity ensued and was a significant theme of the virtual SCES 2020 [105] that was organized by Pascoal Pagliuso, Cris Adriano and Eduardo Miranda in Guaruja, Brazil. But, with this activity came questions. For the order parameter to break TRSB, its two components must belong to different irreducible representations, and thus there must be two intrinsic, bulk superconducting transitions. Experiments at Los Alamos showed that there could be one or two bulk transitions depending on sample-synthesis conditions, that the residual Sommerfeld coefficient was much smaller in single-transition crystals with $T_c$ above 2 K and, in crystals with two transitions, both transitions were suppressed by pressure at the same rate, suggesting the presence of two mesoscale regions in these crystals.[106] Perhaps more concerning, Kerr-rotation experiments on crystals grown by different techniques, having a range of $T_c$'s and only one superconducting transition found no evidence for spontaneous TRSB, but instead a spatially inhomogeneous, field-trainable Kerr response below $T_c$.[107] Though UTe$_2$ may not be a chiral superconductor, its continued study holds potential for clarifying long-open issues on relationships among unconventional superconductivity, magnetic, quantum criticality and now topology.

So, what lies ahead? As the past has shown, surprises happen, but meaningful progress requires more than serendipity. A steady commitment to the discovery of new heavy-fermion materials and their study is essential. This path has proven to be fruitful at Los Alamos and elsewhere. Theoretical modelling, simulation and computation of SCES phenomena have made remarkable progress in the past 40 years; nevertheless, the challenge of handling the many-body physics central to a microscopic understanding of the quantum-entangled degrees-of-freedom in heavy-fermion/Kondo-lattice/Anderson-lattice systems is daunting. Purposefully integrating experiment and theory holds promise and is beginning to yield results. An example is the effort of an experiment/theory team, led out of Los Alamos by Filip Ronning and past group member Marc Janoschek, to understand the Kondo-/Anderson-lattice physics in sufficient first-principles detail that it is allows materials-specific prediction of response functions in related systems. In combination with theory, their charge- and spin-spectroscopy studies on prototypical examples CePd$_3$ and CeIn$_3$ have revealed in new detail how local moments hybridize with itinerant states to form a coherent band of heavy quasiparticles at low temperatures, [108,109] which in the case of CeIn$_3$ is captured quantitatively in a tractable Hamiltonian [109]. Understanding how degrees-of-freedom become quantum entangled as a function of temperature is needed for a more complete picture of the physics, but that understanding first requires a means of measuring entanglement. Allen Scheie, a new addition to the Los Alamos SCES effort, has shown that entanglement and non-locality in insulators can be extracted from high resolution neutron-scattering experiments and theoretical modeling.[110] Extending these techniques to heavy-fermion metals will not be straightforward but should be possible and, when combined with information from the Ronning/Janoschek effort, offers real promise for the future.

## ACKNOWLEDGEMENTS

We are deeply indebted to our colleagues and numerous collaborators who have been an essential part of Los Alamos' SCES history and would be remiss without mentioning Hans Ott and David Pines. Long-term visitors, including Jon Lawrence, Vladimir Sidorov and Hiroshi Yasuoka, have played key roles. Greg Stewart and Cedomir Petrovic generously offered their detailed recollection of special developments in this history. Progress at Los Alamos would not have been possible without the dedication of postdocs with whom we have had the fun of working, and they are the most significant legacy of the SCES effort. Of the well over 80 postdocs, a handful have continued SCES work at Los Alamos, but others have moved on to set their own agenda and become leaders in the SCES community. Finally, we thank the U.S. Department of Energy, Office of Basic Energy Sciences, Division of Materials Science and Engineering for continued support of SCES research.